\documentclass[twocolumn,aps]{revtex4}

\usepackage{graphicx}
\usepackage{amsmath}
\usepackage{amsfonts}
\usepackage{amssymb}

\begin{document}

\title{Subband structure of II-VI modulation-doped magnetic quantum wells}

\author{Henrique J. P. Freire}
 \email{freire@if.sc.usp.br}
\author{J. Carlos Egues}
 \altaffiliation[Present address: ]{University of Basel, Institute of Physics, Klingelbergstrasse 82, CH-4056 Basel, Switzerland}
 \email{egues@if.sc.usp.br}
\affiliation{Departamento de F\'{\i}sica e Inform\'{a}tica, Instituto de F\'{\i}sica de S\~{a}o Carlos, \\
Universidade de S\~{a}o Paulo, C.P.369, 13560-970 S\~{a}o Carlos-SP, Brazil}

\begin{abstract}
Here we investigate the spin-dependent subband structure of newly-developed
Mn-based modulation-doped quantum wells. In the presence of an external
magnetic field, the \textit{s-d} exchange coupling between carriers and
localized \textit{d} electrons of the Mn impurities gives rise to large spin
splittings resulting in a magnetic-field dependent subband structure. Within
the framework of the effective-mass approximation, we self-consistently
calculate the subband structure at zero temperature using Density Functional
Theory (DFT) with \textit{a} Local Spin Density Approximation (LSDA). We
present results for the magnetic-field dependence of the subband structure of
shallow ZnSe/ZnCdMnSe modulation doped quantum wells. Our results show a
significant contribution to the self-consistent potential due to the
exchange-correlation term. These calculations are the first step in the study
of a variety of interesting spin-dependent phenomena, e.g., spin-resolved
transport and many-body effects in polarized two-dimensional electron gases.
\end{abstract}

\maketitle

Recent advances in nanoengineering of Mn-based semiconductor heterostructures
open up an exciting area of research which combines both magnetism and quantum
size effects. The successful achievement of high-doping carrier densities in
Mn-based II-VI layered compounds \cite{Smorchkova96} provides a new and
promising ground for research: spin-polarized two-dimensional magnetotransport
in (semi)magnetic wells. Recently, Smorchkova et al.\ \cite{Smorchkova97}
reported Hall-effect measurements in spin-polarized two-dimensional electron
gases in novel \textit{n}-doped digital-magnetic heterostructures
\cite{Longitudinal,long-theory}.

Here we investigate the magnetic-field dependence of the spin-dependent
electronic structure of a modulation-doped digital magnetic quantum well (QW) 
\cite{Smorchkova97,Awschalon99,Knobel00,Berry00,Quinn00} using Density Functional 
Theory (DFT) with \textit{a} Local Spin Density 
Approximation (LSDA). We consider a Digital Magnetic Heterostructure (DMH)
\cite{Awschalon99} in which magnetic Mn atoms are digitally incorporated in a
planar configuration inside the well. In the presence of an external magnetic
field, the \textit{s-d} exchange coupling between carriers and localized
\textit{d} electrons of the Mn impurities yields large spin splittings that
exceed the inter-Landau level (LL) spacing $\hbar\omega_{c}$ \cite{Furdyna88}.
This gives rise to a magnetic-field dependent subband structure and a net
spin-polarization of the electron gas even at low magnetic fields.%

\begin{figure}[h]
\begin{center}
\includegraphics{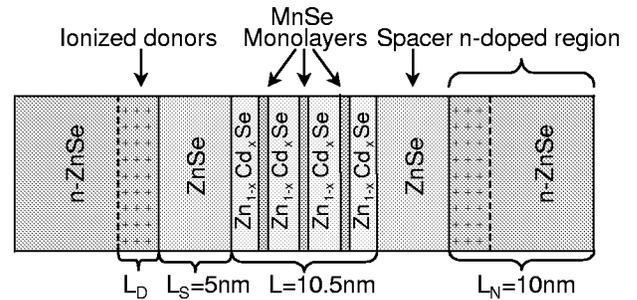}
\caption{Modulation-doped semimagnetic quantum well: the structure consists of
ZnSe layers sandwiching a ZnCdSe layer containing digitally-incorporated MnSe
monolayers. Adjacent to the ZnSe spacer there are \textit{n}-doped regions
(electron reservoirs). Two depleted regions of length $L_{D}$ arise within the
\textit{n}-doped layer as the doping electrons move into the well.}%
\label{figQW}
\end{center}
\end{figure}

\emph{Physical system.} We consider the modulation-doped semi-magnetic quantum
well shown in Fig. 1. The QW structure consists of \textrm{ZnSe} layers
surrounding a 35-monolayer (ML) \textrm{ZnCdSe}/\textrm{MnSe} heterostructure
with six equispaced digitally-incorporated \textrm{MnSe} monolayers, denoted
by $6\times\frac{1}{2}$ ML after Crooker et al. \cite{Crooker95}. This
notation means there are $6$ \textrm{Mn} planes with nominal planar
concentration $x_{p}=1/2$. Adjacent to each \textrm{ZnSe} spacer, there are
\textit{n}-doped regions (impurity concentration $N_{i}=1.2\times10^{17}%
\thinspace\text{cm}%
^{-3}$) that serve as \emph{electron reservoirs}. Electrons from the
\textit{n}-doped regions fill up the confined energy levels of the QW thus
forming a two-dimensional electron gas (2DEG) and leaving behind depleted
regions of length $L_{D}$ in the reservoirs. As the magnetic field is changed,
we expect modifications in the self-consistent potential and in the
corresponding bandstructure due to \textit{i}) magnetic-field dependent
contributions to the confining potential and \textit{ii}) rearrangement of
electrons in the confined levels because of the magnetic-field dependent
Landau-level degeneracy $eB/h$.

\emph{DFT/LSDA approach.} Within Density Functional Theory, we solve the
self-consistent Kohn-Sham equations using \textit{a} Local Spin Density
Approximation scheme to obtain the magnetic field dependent subband structure,
modulation-doped confining potential, and carrier concentration. We consider
the system at zero temperature and perform the calculations within the
framework of the effective-mass approximation. The dominant contribution to
the magnetization comes from the localized moments of the Mn ions. These are
treated as a paramagnetic collection of ions by using an effective Brillouin
function. However, we also consider the effects of the spin density on the
exchange-correlation (XC) term. This is done within the interpolation scheme
of Barth and Hedin \cite{Barth72} as parametrized by Gunnarsson and Lundqvist
\cite{Gunnarsson76}. Since the energy scale in our system (tens of meV) is the
same as that of the calculated XC contribution to the confining potential (up
to 18 meV), we expect XC to significantly affect the self-consistent subband
structure. Thus, to assess the role of exchange and correlation in the
electronic structure in these novel geometries, we contrast our DFT/LSDA
results with similar ones within the Hartree approximation \cite{NoteHartree}.

\emph{Results and discussions.} We present results for \emph{shallow}
ZnSe/ZnCdMnSe modulation-doped QWs. By shallow we mean the confining potential
($25$\thinspace\text{meV}) is small enough so that $\left(  \sim\right)
10.5%
\thinspace\text{nm}%
$ wide wells have only a few confined subbands. For these geometries,
\textit{n}-doped regions with concentrations near $10^{17}%
\thinspace\text{cm}%
^{-3}$ provide carriers that fill up all the QW levels thus serving as a
reservoir of electrons (``reservoir hypothesis'' \cite{Reservoir8194}). We
obtain the magnetic-field dependence of the subband structure and our results
show a significant contribution to the self-consistent potential due to the XC
term. We have calculated the self-consistent potential profile, electronic
structure and the two-dimensional electronic density for magnetic fields
ranging from zero to ten tesla.%

\begin{figure}[h]
\begin{center}
\includegraphics{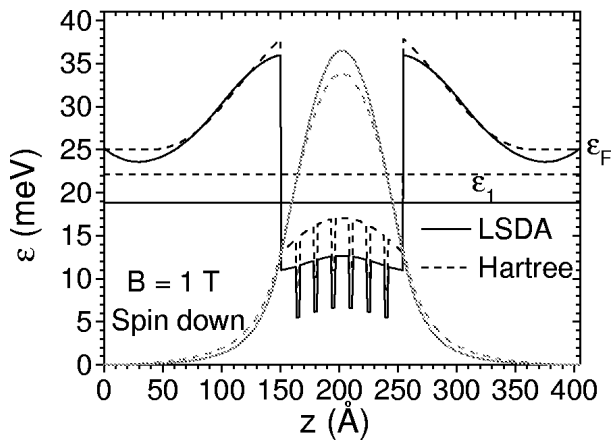}
\caption{Calculated modulation-doped potential profile, subband energy
$\varepsilon_{1}$ and wavefunction squared for spin-down electrons in the
structure shown in Fig. 1 within the Hartree (dashed lines) and the LSDA
(solid lines) approximations. The Fermi energy $\varepsilon_{F}$ is set by the
chemical potential of the reservoir, arbitrarily set to $25\,\mathrm{meV}$.
The main effect of the XC contribution in the LSDA calculation is to lower the
confining potential and the subband energy $\varepsilon_{1}$ for the spin-down
electrons when compared to the Hartree case.}%
\label{figProfile}%
\end{center}
\end{figure}

Figure 2 shows the potential profile, subband energy and wavefunction squared
for spin-down electrons at $B=1%
\thinspace\text{T}%
$. The potential profile consists of the structural confining potential, the
magnetic field and spin-dependent \textit{s-d} exchange energy within the
magnetic monolayers, the Hartree contribution, the Zeeman energy, and the
spin-dependent XC energy. The resulting self-consistent potential has the
usual modulation-doped profile that arises from the addition of the Hartree
energy of the confined electrons to the confining potential. We contrast the
LSDA calculation to the Hartree one. First we note that, despite the
self-consistent Hartree result being fully converged, the LSDA energy profile
is not flat at both ends as one would expect. The reason for this behavior may
be that the LSDA parametrization of Gunnarsson and Lundqvist
\cite{Gunnarsson76} is not appropriate to describe XC at very low electronic
densities such as those within the wavefunction tails [their results are for
Wigner-Seitz radius $r_{s}$ (in units of effective Bohr radius) within the
range $1-9$, while our data has $r_{s}$ from $20$ to $300$ at the tails].
Nevertheless, as most of the wave function is confined inside the well region,
we do not expect these non-flat regions to drastically affect the electronic
structure of the system. In fact, we have tested this hypothesis by abruptly
setting the XC contribution to zero for $r_{s}$ greater than $9$\ and we have
not found any substantial deviation in our results.

\begin{figure}[h]
\begin{center}
\includegraphics{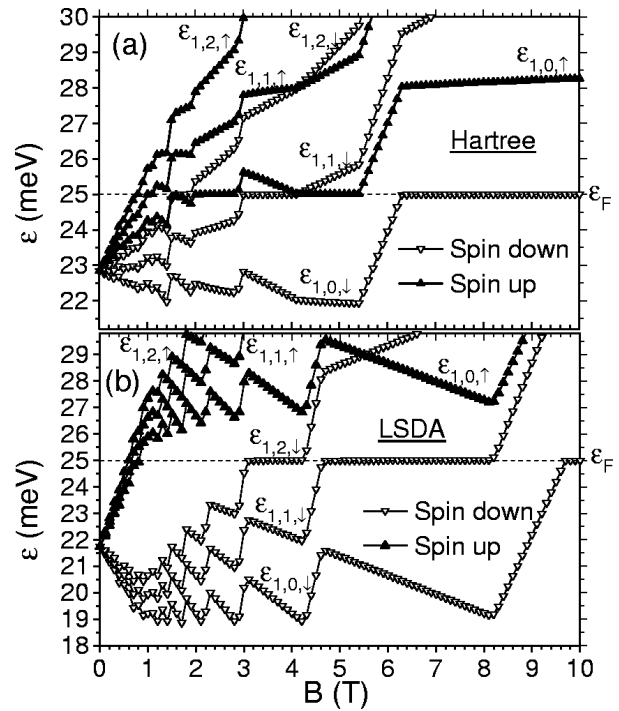}
\caption{Fan diagrams for the first three spin-down (triangle down) and
spin-up (triangle up) Landau levels for both Hartree (a) and LSDA (b)
calculations. Note that the exchange-correlation contribution enhances the
spin-splitting (b) and makes the 2DEG fully polarized at lower magnetic field
($\sim1\thinspace\text{T}$) than in the Hartree case ($\sim3\thinspace\text{T}$).
The Fermi level $\varepsilon_{f}$ (dashed line) remains constant at 25 meV
(electron reservoir) while the Landau levels oscillate as the spin-dependent
potential profile changes with the magnetic field.}%
\label{figLandauLevels}%
\end{center}
\end{figure}

We can clearly see from Fig. 2 that the XC contribution significantly lowers
the energy $\varepsilon_{1}$ for the spin-down electrons. This effect is
larger for the spin-down electrons than for spin-up electrons because the
electron gas is spin-down polarized. Pauli's exclusion principle allows (does
not allow) a spin up (down) electron and the majority spin down electrons to
overlap thus raising (lowering) the electrostatic energy.%

In Fig. 3 we plot the first three spin-resolved Landau-level fan diagrams for
both the Hartree (a) and the LSDA (b) cases. Note that the XC enhances the
spin splitting because it decreases (increases) the spin down (up) energy in
the case of a spin-down polarized gas. The second point is that the LLs are
pinned to the Fermi level $\varepsilon_{F}$ for distinct ranges of magnetic
fields (see plateaus at $\varepsilon_{F}$) \cite{NoteEf}. In the present
model, this follows from the ``reservoir hypothesis'' which allows for two
competing mechanisms to act: \textit{i)} as the magnetic field increases, the
LLs energies also increase and cross the Fermi energy, thus expelling
electrons out of the quantum well; \textit{ii)} on the other hand, less
electrons means less Coulomb repulsion and the energy tends to lower. An
equilibrium between these two mechanisms is responsible for the plateaus. Note
that the pinning of the Landau level in the present model is different from
the standard one used to describe the Integer Quantum Hall Effect (IQHE)
\cite{vonKlitzing80}. There, localized states in broadened Landau levels are
responsible for the Fermi-level pinning \cite{Stormer83}.

One experimental relevant quantity affected by the electronic structure is the
two-dimensional electron density $n_{2D}$. At zero magnetic field we find that
$n_{2D}$ is $2.175\times10^{11}%
\thinspace\text{cm}%
^{-2}$ and $1.444\times10^{11}%
\thinspace\text{cm}%
^{-2}$, respectively, for LSDA and Hartree calculations. This considerable
difference comes about because XC lowers the subband energies as compared to
the Hartree case thus leading to an effective increase in $n_{2D}$.

We have studied the self-consistent electronic structure of a modulation-doped
semimagnetic quantum well including exchange and correlation of electrons
within the DFT/LSDA scheme. We found that for shallow geometries the
exchange-correlation contribution is significant to the electronic structure
as it \textit{i}) lowers the subband energy of the majority spin-down
electrons and \textit{ii}) enhances the spin-splitting thus making the
electron gas fully spin-polarized at lower magnetic fields ($\sim1%
\thinspace\text{T}%
$) than in the Hartree case ($\sim3%
\thinspace\text{T}%
$). Our results are preliminary but do point to the relevance of the XC in
shallow ZnSe semimagnetic QWs, in contrast to GaAs systems. The calculations
reported here are a first step in the study of a variety of interesting
spin-dependent phenomena, e.g., spin-resolved transport and many-body effects
in spin-polarized two-dimensional electron gases.

This work is being supported by FAPESP (99/06868-3), Brazil. We thank Prof.
Nitin Samarth (PennState) for suggesting this problem to us. One of us (HJPF)
acknowledges the hospitality at Prof. Samarth's group during his visit.


\begin{thebibliography}{99}
\bibitem{NoteHartree}The DFT/LSDA scheme accounts for XC. The Hartree
approximation is obtained by setting XC to zero in this scheme.
\bibitem{NoteEf}$\varepsilon_{F}$ remains constant at the chemical potential
of the reservoir. Note that because the quantum well is shallow and the donor
concentration within the \textit{n}-doped regions is high, this region is
never completely depleted.
\end{thebibliography}
\end{document}